\documentclass{svmult}

\usepackage{mathptmx}       
\usepackage{helvet}         
\usepackage{courier}        
\usepackage{makeidx}         
\usepackage{graphicx}        
\usepackage{multicol}        
\usepackage[bottom]{footmisc}

\makeindex             

\newcommand{\Nc}{N_{\rm c}}
\newcommand{\Nf}{N_{\rm f}}
\newcommand{\LQCD}{\Lambda_{\rm QCD}}

\begin{document}

\title*{Views of the Chiral Magnetic Effect}
\author{Kenji Fukushima}
\institute{Kenji Fukushima \at Department of Physics,
           Keio University, 3-14-1 Hiyoshi, Kohoku-ku, Yokohama-shi,
           Kanagawa 223-8522, JAPAN,
           \email{fuku@rk.phys.keio.ac.jp}}
\maketitle

\abstract{
My personal views of the Chiral Magnetic Effect are presented, which
starts with a story about how we came up with the electric-current
formula and continues to unsettled subtleties in the formula.  There
are desirable features in the formula of the Chiral Magnetic Effect
but some considerations would lead us to even more questions than
elucidations.  The interpretation of the produced current is indeed
very non-trivial and it involves a lot of confusions that have not
been resolved.
}


\section{Introduction -- Discovery of the Chiral Magnetic Effect}
\label{sec:discovery}

The Chiral Magnetic Effect (CME) is concisely summarized in the
following handy formula;
\begin{equation}
 \vec{j} = \Nc \sum_f \frac{q_f^2\mu_5}{2\pi^2}\vec{B} \;,
\label{eq:cme}
\end{equation}
which represents an electric current associated with the non-zero
chirality and the external magnetic field $\vec{B}$.  Here $\Nc$
stands for the number of colors in quantum chromodynamics (QCD) and
$q_f$ represents the electric charge carried by the quark flavor $f$
where $f$ runs over \textit{up}, \textit{down}, \textit{strange},
etc.  Equation~(\ref{eq:cme}) looks simple, but the physical meaning
of this CME current is far from simple.  Let me begin with telling
some historical remarks on the discovery of the CME-current formula,
hoping that it may be instructive and even inspiring to some readers.

When we, Harmen~Warringa, Dima~Kharzeev, and I, started working on the
computation of $\vec{j}$, we had no \textit{a priori} idea about the
final answer, hence we did not really expect that the final result
should be such elegant and beautiful.  For several years Harmen and
Dima had been working on the implication of axion physics in the
context of the relativistic heavy-ion collision [I will come back to
  the relevance of the CME to axion physics later.]  At that time,
around the year of 2007, I was thinking of a different (but related)
physics problem, i.e.\ color-superconducting states in a strong
$B$ inspired by a pioneering work~\cite{Ferrer:2005vd}.  Harmen and I
just chatted in the corridor of the RIKEN BNL Research Center (RBRC)
about $B$-effects on color superconductivity, which was soon promoted
to intriguing discussions, and a fruitful collaboration after all.
Nearly simultaneously with the successful completion of our project on
color superconductivity in $B$~\cite{Fukushima:2007fc} (see also
Ref.~\cite{Noronha:2007wg} for an accidental coincidence of the
research interest with our Ref.~\cite{Fukushima:2007fc}), a monumental
paper by Harmen, Dima, and Larry~McLerran
appeared~\cite{Kharzeev:2007jp}.  While we were finalizing the
color-superconductivity paper (or struggling with referees, probably),
Harmen excitedly explained the idea of the Chiral Magnetic Effect to
me.  Also, I clearly remember that Larry came over mischievously (as
always) to ask about the strength of \textit{my} $B$ in the
neutron-star environment ($eB\sim 10^{15}{\rm gauss}$ at most on the
magnetar surface).  As compared to \textit{their} $B$ produced in the
relativistic heavy-ion collision where $eB\sim 10^{20}{\rm gauss}$
could be reached, mine was only negligible...  Indeed, historically
speaking, the recognition of such $B$ as strong as the QCD energy
scale $\LQCD$ in realistic circumstances was an important turning
point to get the $B$-physics research into gear.  In other words,
physics researches at $eB\sim\LQCD^2$ have come to make pragmatic
sense rather than purely academic one since this turning point in
2007.  There was really a tremendous change in the attitude of
researchers.

One year later, Harmen invited me to his continued project with Dima
on the Chiral Magnetic Effect.  In their first paper the formula was
given in a different style from what is known today, namely, it was
not the current but the charge separation $Q$ expressed
as~\cite{Kharzeev:2007jp}
\begin{equation}
 Q = 2Q_w \sum_f |q_f| \gamma(2|q_f\Phi|) \;.
\label{eq:cme_org}
\end{equation}
Here $Q_w$ is the topological charge (i.e.\ counter part of $\mu_5$ in
Eq.~(\ref{eq:cme})) and $\gamma(x)$ is a function dependent on the
microscopic dynamics of quark matter.  According to the analysis in
Ref.~\cite{Kharzeev:2007jp} one can approximate $\gamma(x)$ by a
simple function; $\gamma(x\le 1) = x$,~ $\gamma(x\ge 1) = 1$.  This
means that, if the magnetic flux per unit topological domain, $\Phi$,
is large enough, $Q\approx 2Q_w\sum_f |q_f|$.  This result is
naturally understood from the index theorem, i.e.\
$2Q_w = N_5 = N_L - N_R$.  Under such strong $B$, all the spin
directions should completely align in parallel with $\vec{B}$, and
thus the momentum directions are uniquely determined in accord with
the chirality.  All produced chirality should contribute to the charge
separation, leading to $Q\approx N_5 \sum_f|q_f|$ that is nothing but
$2Q_W\sum_f |q_f|$.  In the weak field case, on the other hand,
$Q\approx 4\Phi Q_w\sum_f q_f^2$ was the theoretical estimate.

Equation~(\ref{eq:cme_org}) is as a meaningful formula as
Eq.~(\ref{eq:cme}), but the determination of $\gamma(x)$ requires some
assumptions.  Besides, since the formula involves $Q_w$, it is
unavoidable to think of topologically non-trivial gauge
configurations.  As a matter of fact, Harmen and I once tried to
compute $Q$ concretely on top of the real-time topological
configuration, namely, the L\"{u}scher-Schechter classical
solution~\cite{Luscher:1977cw,Schechter:1977qg}, which turned out to
be too complicated to be of any practical use.  Then, Harmen hit on a
brilliant idea to deal with $Q_w$, or strictly speaking, an idea to
skirt around $Q_w$.  [He invented another nice trick later to treat
  $Q_w$ more directly.  I will come to this point later.]  The crucial
point is the following;  it is not the topological charge $Q_w$ but
the chirality $N_5$ that causes the charge separation.  It is tough to
think of $Q_w$, then what about starting with $N_5$ not caring too
much about its microscopic origin?  If one wants to fix a value of
some number, one should introduce a chemical potential conjugate to
the number.  In this case of $N_5$, the necessary ingredient is the
chiral chemical potential $\mu_5$ that couples the chiral-charge
operator $\bar{\psi}\gamma^0\gamma^5\psi$.  In my opinion the
introduction of $\mu_5$ was a simple and great step to make the CME
transparent to everybody.  In this way the CME has eventually gotten
equipped with enough simplicity and clarity.

The remaining task was to answer the following question;  what is
$\vec{j}$ in a system with both $\mu_5$ and $\vec{B}$?  Harmen and I
were first going to calculate the expectation value of the current
operator $\bar{\psi}\gamma^\mu \psi$ directly (see the derivation~A in
Ref.~\cite{Fukushima:2008xe}).  To this end we had to solve the Dirac
equation in the presence of $\mu_5$ and $\vec{B}$ to construct the
propagator.  Now I am very familiar with the way how to do this
explicitly, but when we started working on this project, we had not
had enough expertise yet, apart from some straightforward calculations
in color superconductivity.  Some years later Harmen, Dima, and I
wrote a paper in which we reported the diagrammatic method to derive
Eq.~(\ref{eq:cme}) (see Appendix~A in Ref.~\cite{Fukushima:2009ft}).
Let me briefly explain this derivation here;  the electric current in
the $z$-axis direction is written in terms of the propagator as
\begin{eqnarray}
 j_z &=& \Nc \sum_f \frac{q_f|q_f B|}{2\pi} \sum_n \int^T
  \frac{dp_0}{2\pi} \int \frac{dp_z}{2\pi} \int \frac{dx}{L_x}\,
  \nonumber\\
 &&\qquad\qquad\qquad\qquad \times
  \mathrm{tr}\Bigl[ \gamma^z P_n(x) \frac{i}{{\tilde p}_\mu \gamma^\mu
  + \mu_5\gamma^0\gamma^5 - M_f} P_n(x) \Bigr] \;,
\label{eq:cme_prop}
\end{eqnarray}
where the $p_0$-integration is either at $T=0$ or the Matsubara sum at
$T\neq0$.  If we choose the gauge as $A_0=A_x=A_z=0$ and $A_y=Bx$, the
tilde momentum in the denominator is
${\tilde p}=(p_0,0,-\mathrm{sgn}(qB)\sqrt{2|qB|n},p_z)$.  We do not
need the explicit form of the Landau wave-functions $P_n(x)$ that take
a $4\times 4$ matrix structure in Dirac space.  Because we are
interested in $\vec{j}\parallel \vec{B}$ here, $\gamma^z$ commutes
with $P_n(x)$ and thus we need only $P_n(x)^2$ which equals $1$ for
$n>0$ and $(1+i\mathrm{sgn}(q_f B)\gamma^x\gamma^y)/2$ for $n=0$.
After some calculations one can confirm that Eq.~(\ref{eq:cme_prop})
is reduced to Eq.~(\ref{eq:cme}) regardless of the temperature $T$ and
the flavor-dependent mass $M_f$.  Let me make a comment on this rather
na\"{i}ve calculation.  In most cases the proper-time method is the
best way to proceed in theoretical
calculations~\cite{Schwinger:1951nm,Gusynin:1994xp} and the above form
of the quark propagator is not widely known.  For the purpose of
calculating a finite quantity like the CME current, I would like to
stress that the above quark propagator should be equally useful.
Actually it is almost obvious in Eq.~(\ref{eq:cme_prop}) that any
contributions from the Landau non-zero modes are vanishing and the
current arises from the Landau zero-mode only.

Coming back to the story of our first attempt to discover $\vec{j}$,
I remember that Harmen and I came to the office and brought different
answers every morning and had the hottest discussions all the day.  It
took us a few days until we eventually convinced ourselves to arrive
at the right answer.  Later on, Harmen had great efforts to dig out
several independent derivations of Eq.~(\ref{eq:cme}) while preparing
for our paper.  Among various derivations we first found the one based
on the thermodynamic potential (i.e.\ the derivation~C in
Ref.~\cite{Fukushima:2008xe}).  Because this calculation plays some
role in later discussions on the physical interpretation of the CME
current, let us take a closer look at the detailed derivation using
the thermodynamic potential.

The most essential ingredient is the quasi-particle energy dispersion
relation in the presence of $B$ and $\mu_5$.  For $\vec{B}$ along the
$z$-axis, one can solve the Dirac equation to find the dispersion
relation,
\begin{equation}
 \omega_{p,s}^2 = \Bigl[ (p_z^2 + 2|q_f B|n)^{1/2}
  +\mathrm{sgn}(p_z)\, s \mu_5 \Bigr]^2 + M_f^2 \;,
\end{equation}
where $s$ is the spin, $q_f$ and $M_f$ are the electric charge and the
mass of quark flavor $f$.  Once the one-particle energy is given, one
can immediately write the thermodynamic potential down as
\begin{equation}
 \Omega = \Nc \sum_f \frac{|q_f B|}{2\pi}\sum_{s=\pm}\sum_{n=0}^\infty
  \alpha_{n,s} \int_{-\infty}^\infty \frac{dp_z}{2\pi} \Bigl[ \omega_{p,s}
  +T\sum_{\pm} \ln \bigl( 1+e^{-(\omega_{p,s}\pm \mu)/T} \bigr) \Bigr]
\end{equation}
at finite temperature $T$ and quark chemical potential $\mu$.  The
spin factor, $\alpha_{n,s}$, is defined as
$\alpha_{n,s}=1\; (n>0)$, $\delta_{s+}\; (n=0,q_f B>0)$,
$\delta_{s-}\; (n=0,q_f B<0)$.  This factor is necessary to take care
of the fact that the Landau zero-mode ($n=0$) exists for one spin
state only.  The current $j_z$ is obtained by differentiating $\Omega$
with respect to the vector potential $A_z$.  Because the vector
potential in the matter sector resides only through the covariant
derivative, the following replacement is possible inside of the
$p_z$-integration,
\begin{equation}
 \frac{\partial}{\partial A_z} = q\frac{d}{d p_z} \;.
\end{equation}
The combination of this derivative and the $p_z$-integration ends up
with the surface terms.  It is the characteristic feature of the
quantum anomaly that a finite answer results from the ultraviolet
edges in the momentum integration.  That is, the CME current reads,
\begin{eqnarray}
 j_z &=& \Nc \sum_f \frac{q_f |q_f B|}{2\pi} \sum_{s=\pm}\sum_n
  \alpha_{n,s} \! \int_{-\Lambda}^\Lambda \! \frac{dp_z}{2\pi}
  \frac{d}{dp_z} \Bigl[ \omega_{p,s} \!+\!
  T\sum_{\pm} \ln \bigl( 1 \!+\!
  e^{-(\omega_{p,s}\pm \mu)/T} \bigr) \Bigr] \nonumber\\
 &=& \Nc \sum_f \frac{q_f |q_f B|}{4\pi^2}
  \bigl[ \omega_{p,\pm}(p_z=\Lambda)
  -\omega_{p,\pm}(p_z=-\Lambda) \bigr] \nonumber\\
 &=& \Nc \sum_f \frac{q_f |q_f B|}{4\pi^2} \bigl[ (\Lambda\pm\mu_5)
  -(\Lambda\mp\mu_5) \bigr]
  = \Nc \sum_f \frac{q_f^2 \mu_5}{2\pi^2}B \;.
\label{eq:cme_thermo}
\end{eqnarray}
Here, in the second and the third lines, $\pm$ appears from the
Landau zero-mode allowed by $\alpha_{n,s}$, i.e.\ $\pm$ amounts to
$\mathrm{sgn}(q_f B)$ which cancels the modulus of $|q_f B|$, and the
matter part drops off for infinitely large
$\omega_{p,s}(p_z=\pm\Lambda)$.  It would be just a several-line
calculation to make sure that Eq.~(\ref{eq:cme_prop}) is equivalent
with Eq.~(\ref{eq:cme_thermo}) and they are calculations at the
one-loop level.  It is also a common character of the quantum anomaly
that the one-loop calculation would often give the full quantum
answer.  Although I do not know any explicit check of the higher-order
loop effects, the above method at the one-loop level is my favorite
derivation of Eq.~(\ref{eq:cme}); all the calculation procedures are
so elementary and transparent.


\section{Chiral Separation Effect}

Soon later, Harmen and I found that a very similar topological current
had been discovered in the neutron-star environment, that is, the
axial current associated with the quark chemical potential $\mu$ and
the magnetic field $\vec{B}$~\cite{Metlitski:2005pr},
\begin{equation}
 \vec{j}_5 = \Nc \sum_f \frac{q_f^2 \mu}{2\pi^2} \vec{B} \;.
\label{eq:cme_5}
\end{equation}
This is a chiral dual version of Eq.~(\ref{eq:cme}).  Nowadays people
call Eq.~(\ref{eq:cme_5}) the Chiral Separation Effect (CSE) in
contrast to Eq.~(\ref{eq:cme}) referred to as the Chiral Magnetic
Effect.  When we learned the fact that Eq.~(\ref{eq:cme_5}) had been
known earlier, our excitement got cooled down a bit.  Also, three years
later, we came to know that the CME formula had been discovered
further earlier.  Now there is a consensus in the community that the
CME formula~(\ref{eq:cme}) was first derived by
Alex~Vilenkin~\cite{Vilenkin:1980fu}.  It was an embarrassment for me
to have overlooked his work until he brought our attention to his old
papers.  In fact an equivalent of Eq.~(\ref{eq:cme}) has been
rediscovered over and over
again~\cite{Giovannini:1997gp,Giovannini:1997eg,Alekseev:1998ds} and I
would not be surprised even if Eq.~(\ref{eq:cme}) is still buried in
further unknown works.  [I am not talking about the recent activities
to derive Eq.~(\ref{eq:cme}) from a deeper insight into physics such
as Berry's curvature~\cite{Son:2012wh,Zahed:2012yu}, hydro or kinetic
approaches~\cite{Kalaydzhyan:2011vx,Gao:2012ix,Stephanov:2012ki,Kalaydzhyan:2012ut},
and so on, which really deserve more investigations.]

The derivation of Eq.~(\ref{eq:cme_5}) is worth discussing here.  The
topological effects in quantum electrodynamics (QED) from
$\Nc\times\Nf$ quarks add terms in the action as
\begin{equation}
 \delta S = \int d^4 x\, \theta(x) \biggl[\partial_\mu j_5^\mu(x)
  +\Nc\sum_f\frac{q_f^2}{16\pi^2}
  \epsilon^{\mu\nu\rho\sigma}F_{\mu\nu}(x)\,F_{\rho\sigma}(x) \biggr] \;,
\label{eq:dS}
\end{equation}
associated with an axial rotation by $\theta(x)$.  In this way we see
that the axial current is not conserved but anomalous.  With the
replacement of $A_0=\mu$ and $\epsilon^{0ijk}\partial_j A_k=B^i$, one
can transform this expression using the integration by parts into
\begin{eqnarray}
 \delta S &=& \int d^4 x\, \partial_i \theta(x) \biggl[ -j_5^i(x)
  -\Nc\sum_f\frac{q_f^2}{2\pi^2} \epsilon^{0ijk} A_0(x) \partial_j
  A_k(x) \biggr] \nonumber\\
 &=& \int d^4 x\, \partial_i \theta(x) \biggl[ -j_5^i(x)
  +\Nc\sum_f\frac{q_f^2}{2\pi^2} \mu B^i(x) \biggr] \;,
\label{eq:dS2}
\end{eqnarray}
from which Eq.~(\ref{eq:cme_5}) immediately follows.  This derivation
also tells us that the $B$-induced current in the right-hand side of
Eq.~(\ref{eq:cme_5}) is nothing but a part of the Chern-Simons
current $\sim \epsilon^{\mu\nu\rho\sigma}A_\nu\partial_\rho A_\sigma$
in QED.\ \ It should be mentioned that the derivation presented above
is a little bit cooked up by me for the illustration purpose and one
should refer to the original paper~\cite{Metlitski:2005pr} for more
careful treatments of the surface integral.

Before going on our discussions, let me point out that the above
derivation implicitly assumes massless quarks.  If quarks are massive,
Eq.~(\ref{eq:dS}) should be modified with an additional term
$2i M_f\langle\bar{\psi}_f \gamma^5\psi_f\rangle$.  This modification
would be harmless as long as the pseudo-scalar condensate is
vanishing, but in principle, Eq.~(\ref{eq:cme_5}) could be dependent
on $M_f$ unlike Eq.~(\ref{eq:cme}) as argued explicitly in
Ref.~\cite{Metlitski:2005pr}.  In fact it is quite subtle whether
Eq.~(\ref{eq:cme_5}) is sensitive to $M_f$ or not, and I will address
this question in an explicit way soon later.

It would be an interesting question how to derive Eq.~(\ref{eq:cme_5})
microscopically just like the ways addressed in the previous section.
In fact I have once tried to prove Eq.~(\ref{eq:cme_5}) based on the
thermodynamic potential by inserting an axial gauge field.  There
must be a way along this line, but I could not solve it (or I would
say that I did not have enough time to find it out...).  Instead,
here, let me introduce another derivation based on the propagator as
in Eq.~(\ref{eq:cme_prop}).

The axial current is expressed as
\begin{eqnarray}
 j_z^A &=& \Nc \sum_f \frac{q_f |q_f B|}{2\pi} \sum_n \int^T
  \frac{dp_0}{2\pi} \int \frac{dp_z}{2\pi} \int \frac{dx}{L_x}\,
  \nonumber\\
 &&\qquad\qquad\qquad\qquad \times
  \mathrm{tr}\Bigl[ \gamma^z\gamma^5 P_n(x) \frac{i}{{\tilde p}_\mu
  \gamma^\mu + \mu\gamma^0 - M_f} P_n(x) \Bigr]
\end{eqnarray}
at finite quark chemical potential $\mu$.  It is easy to see that any
contributions from $n\neq0$ vanish due to the Dirac trace.  Only the
Landau zero-mode produces a term involving $\gamma^x\gamma^y$ which
makes
$\mathrm{tr}(\gamma^0\gamma^x\gamma^y\gamma^z\gamma^5)=-4i\neq0$.
Then, the above expression simplifies as
\begin{eqnarray}
 j_z^A &=& - \Nc \sum_f \frac{q_f^2 B}{2\pi} \int^T
  \frac{dp_0}{2\pi} \int \frac{dp_z}{2\pi} \,
  \mathrm{tr} \Bigl[ \gamma^z \gamma^5 \frac{{\tilde p}_\mu \gamma^\mu
  +\mu\gamma^0 + M_f}{(p_0+\mu)^2 - p_z^2 - M_f^2}
   \gamma^x \gamma^y \Bigr] \nonumber\\
 &=& 4i \Nc \sum_f \frac{q_f^2 B}{2\pi} \int^T \frac{dp_0}{2\pi}
  \int \frac{dp_z}{2\pi} \,
  \frac{p_0 + \mu}{(p_0 + \mu)^2 - p_z^2 - M_f^2} \nonumber\\
 &=& \Nc \sum_f \frac{q_f^2 B}{2\pi}
  \frac{\partial Z(\mu)}{\partial\mu} \;,
\label{eq:current_prop_a}
\end{eqnarray}
where $Z(\mu)$ denotes the partition function at finite density in
(1+1)-dimensional theory (as a result of the dimensional reduction
with the Landau zero-mode), and thus the $\mu$-derivative leads to the
quark density $n$.  In the second line we used
$2(p_0+\mu)/[(p_0+\mu)^2-p_z^2-M_f^2]=(\partial/\partial\mu)
\ln[(p_0+\mu)^2-p_z^2-M_f^2]$.  One might have thought that it is a
simple exercise to evaluate $Z(\mu)$ with the (1+1)-dimensional
integration.  The fact is, however, that the finite-$\mu$ system in
(1+1) dimensions is by no means simple.

In Ref.~\cite{Metlitski:2005pr} one can find exactly the same
expression as above in a slightly different calculation and the
density is written as (see Eq.~(37) in Ref.~\cite{Metlitski:2005pr}),
\begin{equation}
 n_f(T,\mu) = \int \frac{dp_z}{2\pi} \biggl[
  \frac{1}{e^{(\omega_f-\mu)/T}+1} - \frac{1}{e^{(\omega_f+\mu)/T}+1}
  \biggr]
\label{eq:density_naive}
\end{equation}
with $\omega_f=\sqrt{p_z^2+M_f^2}$.  This result is certainly
$M_f$-dependent as suggested in the paragraph below
Eq.~(\ref{eq:dS2}), and this would make a sharp contrast to the
CME current~(\ref{eq:cme}).

We know, however, that the density in the (1+1)-dimensional fermionic
theory arises from the anomaly~\cite{Schon:2000qy} and the
density~(\ref{eq:density_naive}) is not the right answer.  In fact, in
view of the second line of Eq.~(\ref{eq:current_prop_a}), it seems at
a glance that the $\mu$-dependence could be absorbed in the
$p_0$-integration, which already gives us an impression that something
non-natural should be happening.  To see this, let us take one-step
back to the microscopic expression, i.e., the (1+1)-dimensional
partition function reads,
\begin{eqnarray}
 Z &=& 2 i \int^T \! \frac{dp_0}{2\pi}\int \frac{dp_z}{2\pi} \,
  \ln\bigl[ (p_0+\mu)^2-p_z^2-M_f^2 \bigr] \nonumber\\
 &=& i \int^T \! \frac{dp_0}{2\pi}\int \frac{dp_z}{2\pi} \,
  \mathrm{tr}\bigl[\gamma^0(p_0+\mu)-\gamma^z p_z -M_f \bigr] \;,
\end{eqnarray}
from which the $\mu$-dependence could be eliminated by the chiral
rotation (for the zero-mode basis only),
\begin{equation}
 \psi_0 = e^{i\gamma^z\gamma^0 \mu z} \psi'_0 \;,
\label{eq:rotation}
\end{equation}
leading to (here, we shall show results at $T=0$ for simplicity, but
nothing is changed even at finite $T$),
\begin{eqnarray}
 Z &=& i \int \! \frac{dp_0}{2\pi}\int \frac{dp_z}{2\pi} \,
  \mathrm{tr}[e^{i\gamma^z\gamma^0 \mu z}
  (\gamma^0(i\partial_0+\mu)-\gamma^z i\partial_z -M_f)
  e^{i\gamma^z\gamma^0 \mu z}] \nonumber\\
 &=& i \int \! \frac{dp_0}{2\pi}\int \frac{dp_z}{2\pi} \,
  \mathrm{tr}\bigl[ \gamma^0 i\partial_0 -\gamma^z i\partial_z
  - \tilde{M}_f \bigr] \nonumber\\
 &=& \int_{-\Lambda+\mu}^{\Lambda-\mu} \!\frac{dp_z}{2\pi} \,
  \frac{1}{2}\tilde{\omega}_f
  +\int_{-\Lambda-\mu}^{\Lambda+\mu} \!\frac{dp_z}{2\pi} \,
  \frac{1}{2}\tilde{\omega}_f
\end{eqnarray}
with $\tilde{\omega}_f = \sqrt{p_z^2 + |\tilde{M}_f|^2}$, where
$\tilde{M}_f = M_f e^{2i\gamma^z\gamma^0\mu z}$ is the chirally tilted
mass.  The momentum integration is shifted according to the chiral
rotation~(\ref{eq:rotation}).  The first (second) integral corresponds
to the particle (anti-particle, respectively) contribution.  Thus, one
can extract the $\mu$-dependent piece from the surface terms as
follows;
\begin{eqnarray}
 Z &=& \int_{-\Lambda}^\Lambda \frac{dp_z}{2\pi}\tilde{\omega}_f
  + \biggl( \int_{\Lambda}^{\Lambda+\mu}
  +\int_{\Lambda}^{\Lambda-\mu} \biggr) \frac{dp_z}{2\pi}
  \,\tilde{\omega}_f \nonumber\\
 &=& \mu^2 \frac{d^2}{dx^2} \int_{\Lambda}^{\Lambda+x}
  \frac{dp_z}{2\pi} \, \tilde{\omega}_f
  + \mbox{($\mu$-independent terms)} \nonumber\\
 &=& \frac{\mu^2}{2\pi} + \mbox{($\mu$-independent terms)} \;,
\label{eq:1d_density}
\end{eqnarray}
which results in the density $n_f=\mu/\pi$ that is independent of
$M_f$~\cite{Schon:2000qy}.  It is clear from the second line of the
above calculation that the density originates from the ultraviolet
edges, which is the characteristic feature of the anomaly.  The full
quantum answer is then given as
\begin{equation}
 n_f = \frac{\partial Z(\mu)}{\partial\mu} = \frac{\mu}{\pi}
  \quad\Rightarrow\quad
  j_z^A = \Nc\sum_f \frac{q_f^2 \mu}{2\pi^2} B \;.
\label{eq:density}
\end{equation}
Equivalently, if one is interested in deriving the same answer from
Eq.~(\ref{eq:current_prop_a}) directly, one should split the composite
operator as $\bar{\psi}(x)\gamma^z\gamma^5\psi(x)
\to\bar{\psi}(x+\epsilon)\gamma^z\gamma^5\psi(x)$ and insert the
infinitesimal gauge connection from $x$ to $x+\epsilon$.
Interestingly, contrary to Ref.~\cite{Metlitski:2005pr}, the Chiral
Separation Effect~(\ref{eq:cme_5}) is presumably insensitive to the
quark mass just like the Chiral Magnetic Effect~(\ref{eq:cme}).
Whether Eq.~(\ref{eq:cme_5}) is robust or not regardless of $M_f$ is
an important question particularly in the context of the Chiral
Magnetic Wave (CMW)~\cite{Kharzeev:2010gd}.  The anomalous nature of
the density~(\ref{eq:density}) implies that the CMW can exist also in
the chiral-symmetry breaking phase where quarks acquire substantial
mass dynamically.

I would not insist that I could prove the non-renormalization of
Eq.~(\ref{eq:cme_5}) since the above is just a one-loop perturbation
and non-perturbative interactions may change the story;  I would like
to thank Igor~Shovkovy for raising this unanswerable but unforgettable
question.  The interested readers may consult
Refs.~\cite{Gorbar:2009bm,Fukushima:2010zza} for some examples of
non-non-renormalization.  Anyway, I can at least say with confidence
that, if $B$ is super-strong, the reduction to the (1+1)-dimensional
system should be strict, and then Eq.~(\ref{eq:density_naive}) must be
altered, conceivably as $n_f=\mu/\pi$~\cite{Fukushima:2011jc}.

Although the interpretation of Eq.~(\ref{eq:density}) may swerve a bit
from our main stream, I would emphasize that Eq.~(\ref{eq:density}) is
extremely interesting and it would be definitely worth revisiting its
profound meaning.  Actually, Eq.~(\ref{eq:rotation}) has an impact on
the structure of the QCD vacuum.  Let us consider the hadronic phase
with spontaneous breakdown of chiral symmetry.  After the
rotation~(\ref{eq:rotation}), apart from the anomalous term
$\mu^2/(2\pi)$, the system is reduced to that at zero density, which
means that $\chi=\langle\bar{\psi}'_0 \psi'_0 \rangle$ should take a
finite value.  Therefore, in terms of the original fields $\psi_0$,
the chiral condensates form a spiral structure,
\begin{equation}
 \langle\bar{\psi}_0 \psi_0 \rangle = \chi\cos(2\mu z)\;,\quad
 \langle\bar{\psi}_0 \gamma^z\gamma^0\psi_0 \rangle
  = \chi\sin(2\mu z)\;,
\label{eq:spiral}
\end{equation}
which is called the chiral spiral or the dual chiral-density wave (if
$\gamma^5$ is involved)~\cite{Deryagin:1992rw,Nakano:2004cd}.  In
particular, if the above type of the inhomogeneous ground state is
caused by $B$, it is sometimes called the chiral magnetic
spiral~\cite{Basar:2010zd}.

I should emphasize that Eq.~(\ref{eq:density}) does not really require
the dimensional reduction, while the chiral magnetic spiral needs the
pseudo (1+1)-dimensional nature under sufficiently strong $B$.  This
point might be a bit puzzling.  As long as $j_z^A$ is concerned, only
the Landau zero-mode remains non-vanishingly for any $B$, and the
momentum integration is purely (1+1)-dimensional.  The chiral
condensate is, however, not spiral but homogeneous for small $B$
because of contributions from all non-zero Landau levels.  That is,
the genuine chiral condensate is
$\langle\bar{\psi}\psi\rangle = \sum_n \langle\bar{\psi}_n \psi_n\rangle$,
among which only the Landau zero-mode has a special structure as in
Eq.~(\ref{eq:spiral}).  I would conjecture, hence, that there is no
sharp phase transition from the homogeneous chiral condensate at $B=0$
to the chiral magnetic spiral at $B\neq0$, but it may be possible that
the inhomogeneous zero-mode contribution gradually develops, which
exhibits a smooth crossover to the chiral spirals with increasing $B$.


\section{What is the chiral chemical potential?}

Equation~(\ref{eq:cme_5}) is very similar to the CME
current~(\ref{eq:cme}), so that one might have thought at a first
glance that Eq.~(\ref{eq:cme}) emerges trivially from the insertion of
$\gamma^5$ in both sides of Eq.~(\ref{eq:cme_5}).  The relation
between Eqs.~(\ref{eq:cme}) and (\ref{eq:cme_5}) is not such simple,
though.  As a matter of fact, this point was a major source of
confusions about the validity of Eq.~(\ref{eq:cme}).  One can readily
extend the field-theoretical derivation of Eq.~(\ref{eq:cme_5}) using
Eq.~(\ref{eq:dS}) in order to obtain Eq.~(\ref{eq:cme}) by introducing
the axial vector fields $A_\mu^5$, or the chiral gauge fields,
$A_R=(A_\mu+A_\mu^5)/2$ and $A_L=(A_\mu-A_\mu^5)/2$.  Then, in the
same manner as in the previous section, one can formulate the
counterpart of Eq.~(\ref{eq:dS}) associated with a vector rotation by
$\beta(x)$, that is,
\begin{eqnarray}
 \delta S &=& \int d^4 x\, \beta(x) \biggl[\partial_\mu j^\mu(x)
  +\Nc\sum_f\frac{q_f^2}{16\pi^2}
  \epsilon^{\mu\nu\rho\sigma}F^R_{\mu\nu}(x)\,F^R_{\rho\sigma}(x)  \nonumber\\
 &&\qquad\qquad\qquad\qquad\qquad -\Nc\sum_f\frac{q_f^2}{16\pi^2}
  \epsilon^{\mu\nu\rho\sigma}F^L_{\mu\nu}(x)\,F^L_{\rho\sigma}(x)
  \biggr] \;.
\label{eq:dSRL}
\end{eqnarray}
This leads to $-j^i - \Nc\sum_f (q_f^2/2\pi^2)\epsilon^{0ijk}
(A^R_0-A^L_0)\partial_j A_k=0$ just as in Eq.~(\ref{eq:dS2}), and this
is nothing but Eq.~(\ref{eq:cme}) after the identification of $A^5_0$
as $\mu_5$ (see the derivation~D in Ref.~\cite{Fukushima:2008xe}).
Although the derivation may look flawless, it triggered suspicious
views of Eq.~(\ref{eq:cme}), which was first addressed by Toni~Rebhan,
Andreas~Schmitt, and Stefan~Stricker using the Sakai-Sugimoto
model~\cite{Rebhan:2009vc}.  [It should be noted that the CME current
had been exactly reproduced in the holographic
models~\cite{Yee:2009vw}.]

Obviously, one has to deal with the chiral gauge theory with both
$A_R$ and $A_L$ to introduce $\mu_5$ in the above way, and it is
well-known that the anomaly in the chiral gauge theory has a more
complicated structure than that in the vector gauge theory.  Roughly
speaking, the anomaly is a consequence from the inconsistency between
chiral invariance and gauge symmetry.  In the vector gauge theory,
usually, the vector current is strictly conserved due to adherence to
gauge symmetry, and the anomaly is seen in the axial vector channel
only (see Eq.~(\ref{eq:dS})).  In the case in the chiral gauge theory,
however, there is no such strict demand from the theory and it should
be prescription dependent how the anomaly may appear in the vector and
the axial vector currents.  Indeed we can clearly see from
Eq.~(\ref{eq:dSRL}) that the vector current is also anomalous.  There
are two representative results known as the covariant anomaly and the
consistent anomaly, and they can coincide only when the anomaly
cancellation holds, as is the case in the Standard Model.  The authors
of Ref.~\cite{Rebhan:2009vc} claimed that the vector current should be
free from the anomaly and the theory should accommodate the Bardeen
counter-terms to cancel the anomalous terms in Eq.~(\ref{eq:dSRL}).
Then, needless to say, the CME current is vanishing!

This argument scared Harmen and me very much.  In 2009 when
Ref.~\cite{Rebhan:2009vc} came out, Harmen was a postdoc in Frankfurt
and I was also there as a visitor.  Harmen's face is always very
white, but he got even more whity, and we had a lot of discussions on
Ref.~\cite{Rebhan:2009vc} in Frankfurt with a fear that we might have
made a big steaming mistake...  At that time, neither Harmen nor I was
100\% confident in Eq.~(\ref{eq:cme}) (maybe Dima was?), and the
necessity of the Bardeen counter-terms sounded plausible.  This puzzle
was one of the issues discussed in a RBRC workshop, ``P- and CP-odd
Effects in Hot and Dense Matter'' in May, 2010.  One of the invited
participants, Valery Rubakov, wrote a note to clarify this issue based
on the discussions in the workshop~\cite{Rubakov:2010qi}.  The essence
in his argument is the following.  If one introduces $\mu_5$ as the
zeroth component of the axial gauge field, the CME current is gone
indeed.  However, QCD and QED are not the chiral gauge theory.  One
should then introduce $\mu_5$ in a different way as a conjugate to the
Chern-Simons charge.  Therefore, instead of adding a term
$\mu_5 \bar{\psi}\gamma^0\gamma^5\psi$ in a form of the covariant
derivative in the Lagrangian, one should think of the Chern-Simons
current $K^\mu$ which is deduced from
\begin{equation}
 \Nc\sum_f \frac{q_f^2}{16\pi^2} \epsilon^{\mu\nu\rho\sigma}
  F_{\mu\nu}(x)F_{\rho\sigma}(x)
 = \partial_\mu \biggl[ \Nc\sum_f \frac{q_f^2}{4\pi^2}
  \epsilon^{\mu\nu\rho\sigma} A_\nu(x)\partial_\rho A_\sigma(x)
  \biggr] = \partial_\mu K^\mu(x)
\end{equation}
in the QED sector.  The term to be added in the Lagrangian is,
\begin{equation}
 S_{\rm cs} = -\int d^4 x\,\mu_5 K^0(x) =
  -\Nc\sum_f \frac{q_f^2 \mu_5}{4\pi^2} \int d^4 x\,
  \epsilon^{0ijk} A_i(x)\partial_j A_k(x) \;,
\label{eq:action_cs}
\end{equation}
from which Eq.~(\ref{eq:cme}) immediately follows as a result of the
derivative, $j^i = \delta S_{\rm cs}/\delta A_i(x)$.  One may worry
about gauge invariance in the above prescription.  It would be then
more convenient to rewrite $S_{\rm cs}$ in the following way after the
integration by parts, that is manifestly gauge invariant,
\begin{equation}
 S_{\rm cs} = \int d^4 x\, \theta(t) \Nc\sum_f \frac{q_f^2}{16\pi^2}
  \epsilon^{\mu\nu\rho\sigma}F_{\mu\nu}(x)F_{\rho\sigma}(x) \;,
\end{equation}
where $\partial_0 \theta(t) = \mu_5$.  In other words, we can say that
$\mu_5$ is the time derivative of the $\theta$ angle in the QED
sector, which was pointed out already in
Ref.~\cite{Fukushima:2008xe} and the idea of the charge separation
driven by inhomogeneous $\theta$ can be traced back to
Ref.~\cite{Kharzeev:2004ey}.  A subsequent question naturally arises;
what happens if $\theta(x)$ has not only temporal but also spatial
dependence in general?  The Chern-Simons-Maxwell theory with general
$\theta(x)$ provides us with the following modified Maxwell equations;
\begin{eqnarray}
 && \vec{\nabla}\cdot\vec{E} = \rho
  + \Nc\sum_f\frac{q_f^2}{2\pi^2} (\vec{\nabla}\theta)\cdot\vec{B}\;,
\label{eq:cme_charge}\\
 && \vec{\nabla}\times\vec{B} - \partial_0\vec{E} = \vec{j}
  + \Nc\sum_f\frac{q_f^2}{2\pi^2} \Bigl[ (\partial_0\theta)\vec{B}
  - (\vec{\nabla}\theta)\times\vec{E} \Bigr]\;,
\label{eq:cme_current}
\end{eqnarray}
and Faraday's law and Gau\ss's law are not altered.  We see that the
CME current appears in the right-hand side of
Eq.~(\ref{eq:cme_current}) as if it is a part of the external
current.  In this manner we can conclude from
Eq.~(\ref{eq:cme_charge}) that an electric-charge density is induced
by spatially inhomogeneous $\theta(x)$ in the presence of $\vec{B}$.
To the best of my knowledge Eqs.~(\ref{eq:cme_charge}) and
(\ref{eq:cme_current}) are the quickest derivation of the Chiral
Magnetic Effect, as discussed first in Ref.~\cite{Kharzeev:2009fn}.

[After I finished writing this article, I was informed by Toni, one of
the authors of Ref.~\cite{Rebhan:2009vc}, that the confusion about the
CME in the holographic context seems to continue.  I am not able
enough to make any judgment here, and those who want to dive into
this confusion can consult the recent analysis in
Ref.~\cite{BallonBayona:2012wx}.]


\section{What really flows?}
\label{sec:flow}

To tell the truth, I have never gotten any satisfactory answer to the
following question;  what really flows?  I have had various
discussions with people who have various backgrounds, but those
discussions ended up with more confusions than before.  Thanks to
useful conversations, nevertheless, my eyes have been open to various
views of Eq.~(\ref{eq:cme}).  People (including me) say that the CME
current is an \textit{electric current} induced by $\vec{B}$ just like
Ohm's law with the electric field $\vec{E}$.  Let me begin with
suspecting this interpretation that people just take for granted.

In classical electrodynamics Eq.~(\ref{eq:cme_current}) is usually
written in a slightly different way, i.e.,
\begin{equation}
 \vec{\nabla}\times\vec{B} = \vec{j} + \partial_0\vec{E}
  + \Nc\sum_f\frac{q_f^2}{2\pi^2} \Bigl[ (\partial_0\theta)\vec{B}
  - (\vec{\nabla}\theta)\times\vec{E} \Bigr]\;,
\end{equation}
and $\partial_0\vec{E}$ is called the displacement \textit{current}.
We see that the CME current should be a genuine current \textit{if}
$\partial_0\vec{E}$ can be regarded as a real electric current, for
they enter Amp\`{e}re's law on equal footing.  In other words,
\textit{if} the displacement current is not a real current, the CME
current is not, either.  Now, we know from our experience that
$\partial_0\vec{E}$ is only the time derivative of the electric field
and no electric charge flows associated with the displacement
current.  The displacement current certainly plays the equivalent role
as $\vec{j}$ as a source to create $\vec{B}$, but it is clear that
there is no movement of electric charge at all.  It would be therefore
a legitimate claim to insist that the charge separation from the CME
current might be an illusion.  I would emphasize the importance to
distinguish the current and the charge in the argument here.  For
example, the most well-known example of the displacement current is
the problem of the capacitor that is composed of two separate
conductors.  Let a capacitor be connected to the wire with finite
electric current.  More and more electric charge accumulates on the
conductors and produces stronger and stronger electric field inside as
the time goes.  Then, even though two conductors are physically
separate and no electric current flows between them, the displacement
current flows as if the electric current flowed along the wire without
the capacitor.  The distribution of the electric charge stored on the
conductors is, however, totally different depending on the situation
with and without the capacitor.  In this sense, thus, the charge
itself may not flow and the charge separation may not occur with the
CME current also.

A related criticism against the CME current is that the current
computed in Eq.~(\ref{eq:cme_thermo}) for example is the expectation
value of the current operator, $\bar{\psi}\gamma^\mu\psi$, and it is
not necessarily the current.  In fact, there are some studies on the
Chiral Magnetic Effect in the lattice gauge theory;  the correlation
functions of the chirality and the current were measured in
Ref.~\cite{Buividovich:2009wi}, and later Eq.~(\ref{eq:cme}) was
checked directly on the lattice~\cite{Yamamoto:2011gk}.  It is not so
straightforward, however, to interpret these lattice results properly.
A system with a finite electric current could be steady but is out of
equilibrium.  What one can calculate in the thermal system in
equilibrium like the situation of the lattice simulation in Euclidean
space-time is the electric-current conductivity according to the Kubo
formula.  It is a tricky question what
$\langle\bar{\psi}\gamma^\mu\psi\rangle$ really represents in the
lattice simulation.  Let me take one example for concreteness.  If the
system has a condensate of the omega meson, $\omega^\mu$, the
interpolation field of $\omega^\mu$ is $\sim\bar{\psi}\gamma^\mu\psi$
and then $j^\mu=\langle\bar{\psi}\gamma^\mu\psi\rangle\neq0$, but this
does not necessarily mean that the system has a persistent current.
To make this point clearer, the spin operator in terms of the Dirac
matrices is $\hat{S}^z=\frac{i}{4}[\gamma^x,\gamma^y]
=\frac{1}{2}{\rm diag}[\sigma^3,\sigma^3]$, so that the spin
expectation value is
$S^z = \langle\bar{\psi}\hat{S}^z\psi\rangle
 = \frac{1}{2}\langle\phi^\dagger_R \sigma^3 \phi_L\rangle +
   \frac{1}{2}\langle\phi^\dagger_L \sigma^3 \phi_R\rangle$, while the
current expectation value is
$j^z = \langle\bar{\psi}\gamma^z\psi\rangle
 = \langle\phi^\dagger_R \sigma^3 \phi_R \rangle
 - \langle\phi^\dagger_L \sigma^3 \phi_L \rangle$, where $\phi_L$ and
$\phi_R$ are two-component spinors in the left-handed and right-handed
chirality, respectively.  Here, the similarity between $S^z$ and $j^z$
implies that we can regard $j^z$ as a static quantity like the spin
$S^z$, which may well be the most appropriate interpretation of the
lattice measurement.

From the point of view of the theoretical treatment of the electric
current, the formulation based on the linear response theory must be a
good starting point.  I believe that the work along this line in
Ref.~\cite{Kharzeev:2009pj} should be one of the most important
literature to think of physics of the Chiral Magnetic Effect.  They
computed the one-loop diagram on top of the $\mu_5$ background to find
the chiral magnetic conductivity $\sigma_\chi(\omega,p)$.  The result
is consistent with Eq.~(\ref{eq:cme}) in a particular limit;
$\sigma_\chi(\omega=0,p\to0)=\lim_{p\to0}\sigma_\chi(0,p)=
\Nc\sum_f(q_f^2\mu_5/2\pi^2)$ which correctly reproduces the CME
current.  In view of the result of Ref.~\cite{Kharzeev:2009pj}, on the
other hand, it seems $\sigma_\chi(\omega\to0,p=0)=0$.  [This latter
  limit is not manifestly addressed in Ref.~\cite{Kharzeev:2009pj},
  but it is pointed out that the conductivity drops to one third just
  away from $\omega=0$.  It seems to be vanishing from Eq.~(38) and
  Fig.~1 of Ref.~\cite{Kharzeev:2009pj}.]  This is a problem because
the latter limit rather than the former one is more relevant to the
real-time dynamics.  The fact that the former limit ($\omega=0$ first
and $p=0$ next) gives the CME current~(\ref{eq:cme}) suggests that the
CME current should be a static quantity just as measured in the
lattice simulation and thus not a genuine electric current!?

One may still consider that the intuitive argument leading to
Eq.~(\ref{eq:cme_org}) should work anyway.  My impression is also that
all above-mentioned problems are just on the conceptual level (though
I have no idea how to reconcile them) and in practice the CME current
flows according to Eq.~(\ref{eq:cme}) after all.  Indeed if there are
almost massless quarks in a quark-gluon plasma and a strong $\vec{B}$
is imposed on a topological domain, an electric current must be
induced for sure.  An example of the real-time calculation of the CME
current with not $\mu_5$ but a topological domain is quite
instructive in this sense~\cite{Fukushima:2010vw}.  The central
innovation in Harmen's idea (as discussed in
Ref.~\cite{Fukushima:2010vw}) was to mimic the topological domain by
putting $\vec{E}$ and $\vec{B}$ parallel to each other, with which
$\epsilon^{\mu\nu\rho\sigma}F_{\mu\nu}F_{\rho\sigma}\neq0$.  Then, the
particle production occurs via the Schwinger mechanism and the
produced particles are accelerated by the fields, and the electric
current is generated.  The current is time dependent and the
current-generation rate can be analytically written down.  In this
setup the physical origin of the CME current is crystal-clear!  So, if
anything is fishy in physics of the Chiral Magnetic Effect, it should
have something to do with technical defects of $\mu_5$ in equilibrium
circumstances.

Supposing that physics of the Chiral Magnetic Effect should be robust,
let us admit the CME current~(\ref{eq:cme}) as it is to proceed to the
next question, that is actually the central question in this section;
what really flows?

An intuitive explanation tells us that quarks simply flow in a
quark-gluon plasma.  It is, however, based on a classical picture, and
such a picture misses quantum character that is indispensable for
phenomena related to the quantum anomaly.  Look at the derivation of
the CME current in Eq.~(\ref{eq:cme_thermo}).  If this derivation
captures the underlying physics of the CME current, the origin of the
current comes from quarks with infinitely large momenta.  Where are
such fast-moving quarks in the real quark-gluon plasma?  They may
spill out from the vacuum through quantum processes, but how is it
possible to retrieve particles with infinite momenta?  Usually the
quantum anomaly involves ultraviolet regions of the momentum
integration as a loop of virtual particles, meanwhile ultraviolet
particles directly participate in the physical observable in the CME
problem.  It is very hard (at least for me) to imagine that the
current generation in such a way really happens in a physical plasma.
This deliberation brings me a further doubt about the static
evaluation of the CME current.

A natural extension of this question about the origin of the CME
current is whether it exists in the hadronic phase and, if it does,
how the current appears in terms of hadronic degrees of freedom.
Actually this question has been something in mind for a long time
since when we published Ref.~\cite{Fukushima:2008xe}.  In the hadronic
phase an electric current should be attributed to charged pions, but
pions are insensitive to chirality and thus $\mu_5$ or the strong
$\theta$ angle.  One possible answer would be that there is no CME in
the hadronic phase, and if so, it would be fantastic;  the CME current
can be a signature for quark deconfinement, as implied in
Ref.~\cite{Fukushima:2008xe}.  I had heard that Harmen wanted to
analyze the CME using the chiral perturbation theory, though he never
worked it out.

Recently I have clarified what would happen in the hadronic phase and
wrote a paper with one of my students,
Kazuya~Mameda~\cite{Fukushima:2012fg}.  Our conclusion was a very
natural one, and a very perplexing one at the same time.

The CME current is unchanged even in the hadronic phase, which is very
natural since the CME current has the anomalous origin that arises
from ultraviolet fluctuations.  At low energies the anomaly should be
saturated by infrared degrees of freedom, which is sometimes referred
to as the anomaly matching.  This idea is formulated as the
Wess-Zumino-Witten action and the current should be given by the
derivative of the total effective action with respect to the gauge
field.  In this way we found that the leading-order term in the chiral
Lagrangian leads to the current,
\begin{equation}
 j_\chi^\mu = -i \frac{e f_\pi^2}{4} {\rm tr}\Bigl[
  \bigl( \Sigma^\mu - \tilde{\Sigma}^\mu \bigr) \tau^3 \Bigr]
  \simeq e (\pi^- i\partial^\mu \pi^+ - \pi^+ i\partial^\mu \pi^-)
  + \dots \;,
\end{equation}
where $\Sigma^\mu=U^\dagger\partial^\mu U$,
$\tilde{\Sigma}^\mu=(\partial^\mu U)U^\dagger$, and
$U=e^{i\pi^a\tau^a/f_\pi}$ are the standard notation in the chiral
Lagrangian.  The physical meaning of the above expression is plain as
seen from the expansion in terms of the pion fields.  It is a common
form of the probability flow in Quantum Mechanism representing the
electric current associated with the flow of the charged pions.

A more non-trivial contribution comes from the Wess-Zumino-Witten
part, which leads to the current associated with the $\pi^0$
domain-wall~\cite{Son:2007ny}, i.e.,
\begin{equation}
 j_{\rm WZW}^\mu = \Nc\sum_f \frac{q_f}{8\pi^2 f_\pi}
  \epsilon^{\mu\nu\rho\sigma} (\partial_\nu\pi^0)F_{\rho\sigma}\;.
\label{eq:cme_pion}
\end{equation}
This current is very similar to the CME current~(\ref{eq:cme}) and
$\theta(x)$ is just replaced by $\pi^0(x)/(4\pi^2 f_\pi)$.  Although
Eq.~(\ref{eq:cme_pion}) is not the Chiral Magnetic Effect, it would
give us a clue to think about the physical meaning of the CME
current.  Finally, the CME current appears from the so-called contact
part of the Wess-Zumino-Witten action~\cite{Kaiser:2000ck};
\begin{equation}
 S_{\rm P} = \Nc\sum_f \frac{q_f^2}{8\Nf\pi^2}
  \epsilon^{\mu\nu\rho\sigma} \int d^4 x \,
  A_\mu(x) (\partial_\nu A_\rho(x))
  \partial_\sigma \theta(x) \;,
\label{eq:wzw_contact}
\end{equation}
which is just equivalent to the Chern-Simons action already discussed
in Eq.~(\ref{eq:action_cs}).  [$\theta$ in Eq.~(\ref{eq:wzw_contact})
  has a different normalization by $2\Nf$ by convention.]  Naturally
the current derived from Eq.~(\ref{eq:wzw_contact}) should reproduce
the CME current~(\ref{eq:cme}).  This is how one can get the CME
current in the hadronic phase and my surprise lies in the fact that
the pion dynamics is completely decoupled from the CME current.

Because the $\pi^0$ domain-wall looks a bit more intuitive than the
mystical $\theta$ angle, we shall consider a possible interpretation
of the current~(\ref{eq:cme_pion}).  This is certainly a current, but
no charged pions, $\pi^\pm$, are involved in the formula.  Then, it is
as puzzling in Eq.~(\ref{eq:cme_pion}) how the current can flow and
what really flows.

To answer this question, let me emphasize a very useful analogue of
the Josephson current in superconductivity.  The Josephson junction
consists of superconducting materials and a thin layer of insulator
(S-I-S) or non-superconducting metal (S-N-S) sandwiched by them.
There was a big debate about whether the super-current can flow or not
through the insulating barrier.  Of course, there is no Cooper pair
inside of the insulator, and thereby there is nothing that takes care
of the super-current.  It should have been a natural attitude to get
skeptical about such a current~\cite{PhysRevLett.9.147}.  This
situation, a current without current carriers, is quite reminiscent of
our problem of the CME current or the current accompanied by the
$\pi^0$ domain-wall.  Everyone knows that the Josephson current is the
experimental fact today~\cite{PhysRevLett.10.230}.  For the Josephson
current, the coherence is the most important;  in superconductor the
quantum state is characterized by a wave-function just like a problem
in Quantum Mechanics.  In the QCD case, also, such a coherent state is
realized by the condensation of fields, namely, $\pi^0(x)$ in
Eq.~(\ref{eq:cme_pion}) is to be regarded as a macroscopic
wave-function.  One may then raise a question;  the current of the
$\pi^0$ domain-wall may be okay, but what about the CME current?
There is no coherent field but only $\theta(x)$ that is not a
dynamical field but just a space-time dependent parameter!  This is
perfectly a sensible question.  To answer this, I would say that
$\theta(x)$ could be promoted to the dynamical field without
mentioning on a possibility of axion, at least in the hadronic phase.
If the system has a pseudo-scalar (and iso-scalar) condensate such as
the condensate of the $\eta^0$ meson (that forms $\eta'$ with a
mixture with $\eta^3$), it could be mapped to $\theta(x)$ in the chiral
Lagrangian approach.  Once this mapping is noticed, there is no longer
a big conceptual difference between the current in
Eq.~(\ref{eq:cme_pion}) and the CME current in Eq.~(\ref{eq:cme}).
The analogy to the Josephson current may support the reality of the
CME current, but this argument does not tell us anything about the
microscopic constituent of the current yet.

Equation~(\ref{eq:cme_pion}) means that the current can exist just
with the $\pi^0(x)$ profile and the magnetic field, and then the only
possible carrier of the current should be the quark content inside of
$\pi^0$.  Therefore, even in the hadronic phase, I must think that
charged quarks flow to produce the electric current.  Contrary to the
intuition, there is no inconsistency with the notion of quark
confinement.  Regardless of the presence of the flow of quarks, these
flowing quarks can be still confined in a big wave-function of the
$\pi^0(x)$ profile.  In this way, confined quarks can flow without
breaking confinement because of the coherent background of the meson
fields.

In reality it is next to impossible to achieve such an environment
with abundant $\pi^0$ that forms a condensate to test
Eq.~(\ref{eq:cme_pion}) because $\pi^0$ quickly decays into photons
via the anomalous QED process.  This implies that the CME current may
be also diminished by the photon production.  Indeed,
Eq.~(\ref{eq:wzw_contact}) exactly describes such a process of
$\theta(x)$ decaying into $2\gamma$.  It is interesting, besides, that
one $\gamma$ can be provided from $B$ in the case with background
fields.  More specifically, one injected $\gamma$ and another $\gamma$
from $B$ can produce a $\theta$ (or the $\eta^0$ meson), that is
nothing but the Primakoff effect~\cite{Primakoff:1951pj}.  The
Primakoff effect has an application as a tool to detect the
axion~\cite{Sikivie:1983ip,Raffelt:1987im}, which is understandable
from the above argument once $\theta(x)$ is augmented as a dynamical
axion field.  Because physics of the Chiral Magnetic Effect has a
connection to axion physics through $\theta(x)$ (that was actually the
very beginning of the path toward Ref.~\cite{Kharzeev:2007jp}, as I
mentioned), it should be naturally motivated to think of some
application of the Primakoff effect in the context of the Chiral
Magnetic Effect too.  Then, the reverse process of the Primakoff
effect, namely, $\gamma (B) + \theta \to \gamma$, should be the most
relevant to the experimental opportunity.  In the relativistic
heavy-ion collision, the profile of the magnetic field $\vec{B}(x)$
can be estimated by the simulation, and the precise measurement of
$\gamma$ with subtraction of the background from the $\pi^0$ decay is
available nowadays.  The unknown piece in the reverse Primakoff effect
is the profile of the $\theta(x)$ distribution.  Needless to say,
nothing is more important and ambiguous than the concrete distribution
of $\theta(x)$ for any attempt to perform serious computation of the
CME-related phenomena.  This is why most of works on the Chiral
Magnetic Effect address only qualitative predictions.

I think that it must be a very interesting challenge to find a
condensed-matter counterpart in which the Chiral Magnetic Effect may
be visible and testable experimentally.  This is not an unrealistic
desire;  axion physics can be discussed in the so-called topological
magnetic insulator~\cite{Ooguri:2011aa}, and why not the Chiral
Magnetic Effect?  In fact, recently, there are appearing some works
one after another along this line.

\section{My Outlook}

Many people (including me) are still working on the theoretical
aspects of the Chiral Magnetic Effect and its relatives such as the
Chiral Separation Effect, the Chiral Magnetic Wave, etc.  It is highly
demanded to make some firm theoretical estimation about the
experimental observables affected by the CME and related phenomena.
To this end, however, one needs know the early-time dynamics even
before the formation of the quark-gluon plasma.  In fact, the most
interesting regime that has the greatest impact on the CME happens to
be the most difficult regime to describe theoretically.

A missing theoretical link between the coherent wave-function right
after the collision and the thermalized plasma is the last piece of
the jigsaw puzzle.  Particles should be produced from quantum
fluctuations on top of coherent fields (i.e.\ the Color Glass
Condensate; CGC), that translates into the entropy production.  It is
known that the coherent background fields accommodate topological
flux-tubes that play a role of $Q_w$ in Eq.~(\ref{eq:cme_org}).
Produced particles inside of those flux-tubes under a strong $\vec{B}$
should have a characteristic momentum distribution and this would
embody the Chiral Magnetic Effect in a quantitative way.  In my
opinion, thus, the early-thermalization problem must be resolved even
before talking about the observational possibility of the Chiral
Magnetic Effect, or they may have something to do with each other.
This is because, as I stated in the previous section, the real-time
dynamics of the Chiral Magnetic Effect should involve the Schwinger
process of particle production, and the particle production as in the
Lund string model should be responsible for the entropy production
from fields, and thus thermalization ultimately.

One may claim that the complete isotropization and thermal equilibrium
should be no longer required to account for the experimental
observation.  This is true indeed, and this is a good news for the
CME, for the thermalization means that the system should lose any
memory and have only one information, i.e.\ the temperature.  If the
thermalization is incomplete, it would enhance the chance that the
observed distribution of particles may still remember the early-time
environment like the presence of the strong $\vec{B}$ and/or the
topological flux-tubes.  For the purpose of testing the idea as
compared to the experimental data, it is indispensable to perform some
serious simulation of the early-time dynamics of the heavy-ion
collisions.

Unfortunately, there is no successful simulation starting from the CGC
initial condition to achieve some reasonable input for the
hydrodynamics within a reasonable time scale (See
Ref.~\cite{Gale:2012rq} for a latest attempt).  There are so many
theoretical efforts in this direction including
mine~\cite{Fukushima:2011nq} and it should be definitely worth
discussing them, but not here and on another occasion maybe.  In this
article I have discussed physics of the Chiral Magnetic Effect and
presented my views on the physical interpretation.  Now my story has
become a bit too diverging, and I should stop here, with a hope that
some readers may find my views useful for future investigations.

\begin{acknowledgement}
 The author would like to express his sincere thanks to his
 collaborators, Dima~Kharzeev, Kazuya~Mameda, and Harmen~Warringa.
 The contents of this article are based on the fruitful collaboration
 with them.  Especially, this article is dedicated to Harmen~Warringa,
 who inspired me enough to initiate my working on the Chiral Magnetic
 Effect.  Harmen's contribution to physics of the Chiral Magnetic
 Effect should be memorable as long as the CME-related physics is of
 our interest.  The author is also grateful to Tigran~Kalaydzhyan,
 Dima~Kharzeev, Toni~Rebhan, Mikhail~Shaposhnikov, Igor~Shovkovy for
 useful comments on this article.
\end{acknowledgement}
\bibliographystyle{spmpsci}
\bibliography{cme}

\begin{thebibliography}{10}
\providecommand{\url}[1]{{#1}}
\providecommand{\urlprefix}{URL }
\expandafter\ifx\csname urlstyle\endcsname\relax
  \providecommand{\doi}[1]{DOI~\discretionary{}{}{}#1}\else
  \providecommand{\doi}{DOI~\discretionary{}{}{}\begingroup
  \urlstyle{rm}\Url}\fi

\bibitem{Alekseev:1998ds}
Alekseev, A.Y., Cheianov, V.V., Frohlich, J.: {Universality of transport
  properties in equilibrium, Goldstone theorem and chiral anomaly}.
\newblock Phys. Rev. Lett. \textbf{81}, 3503--3506 (1998).
\newblock \doi{10.1103/PhysRevLett.81.3503}

\bibitem{PhysRevLett.10.230}
Anderson, P.W., Rowell, J.M.: Probable observation of the josephson
  superconducting tunneling effect.
\newblock Phys. Rev. Lett. \textbf{10}, 230 (1963).
\newblock \doi{10.1103/PhysRevLett.10.230}.
\newblock \urlprefix\url{http://link.aps.org/doi/10.1103/PhysRevLett.10.230}

\bibitem{BallonBayona:2012wx}
Ballon-Bayona, A., Peeters, K., Zamaklar, M.: {A chiral magnetic spiral in the
  holographic Sakai-Sugimoto model}  (2012)

\bibitem{PhysRevLett.9.147}
Bardeen, J.: Tunneling into superconductors.
\newblock Phys. Rev. Lett. \textbf{9}, 147 (1962).
\newblock \doi{10.1103/PhysRevLett.9.147}.
\newblock \urlprefix\url{http://link.aps.org/doi/10.1103/PhysRevLett.9.147}

\bibitem{Basar:2010zd}
Basar, G., Dunne, G.V., Kharzeev, D.E.: {Chiral magnetic spiral}.
\newblock Phys. Rev. Lett. \textbf{104}, 232,301 (2010).
\newblock \doi{10.1103/PhysRevLett.104.232301}

\bibitem{Buividovich:2009wi}
Buividovich, P., Chernodub, M., Luschevskaya, E., Polikarpov, M.: {Numerical
  evidence of chiral magnetic effect in lattice gauge theory}.
\newblock Phys. Rev. \textbf{D80}, 054,503 (2009).
\newblock \doi{10.1103/PhysRevD.80.054503}

\bibitem{Deryagin:1992rw}
Deryagin, D., Grigoriev, D.Y., Rubakov, V.: {Standing wave ground state in high
  density, zero temperature QCD at large N(c)}.
\newblock Int. J. Mod. Phys. \textbf{A7}, 659--681 (1992).
\newblock \doi{10.1142/S0217751X92000302}

\bibitem{Ferrer:2005vd}
Ferrer, E.J., de~la Incera, V., Manuel, C.: {Magnetic color flavor locking
  phase in high density QCD}.
\newblock Phys. Rev. Lett. \textbf{95}, 152,002 (2005).
\newblock \doi{10.1103/PhysRevLett.95.152002}

\bibitem{Fukushima:2011jc}
Fukushima, K.: {QCD matter in extreme environments}.
\newblock J. Phys. \textbf{G39}, 013,101 (2012).
\newblock \doi{10.1088/0954-3899/39/1/013101}

\bibitem{Fukushima:2011nq}
Fukushima, K., Gelis, F.: {The evolving Glasma}.
\newblock Nucl. Phys. \textbf{A874}, 108--129 (2012).
\newblock \doi{10.1016/j.nuclphysa.2011.11.003}

\bibitem{Fukushima:2008xe}
Fukushima, K., Kharzeev, D.E., Warringa, H.J.: {The Chiral magnetic effect}.
\newblock Phys. Rev. \textbf{D78}, 074,033 (2008).
\newblock \doi{10.1103/PhysRevD.78.074033}

\bibitem{Fukushima:2009ft}
Fukushima, K., Kharzeev, D.E., Warringa, H.J.: {Electric-current susceptibility
  and the chiral magnetic effect}.
\newblock Nucl. Phys. \textbf{A836}, 311--336 (2010).
\newblock \doi{10.1016/j.nuclphysa.2010.02.003}

\bibitem{Fukushima:2010vw}
Fukushima, K., Kharzeev, D.E., Warringa, H.J.: {Real-time dynamics of the
  Chiral Magnetic Effect}.
\newblock Phys. Rev. Lett. \textbf{104}, 212,001 (2010).
\newblock \doi{10.1103/PhysRevLett.104.212001}

\bibitem{Fukushima:2012fg}
Fukushima, K., Mameda, K.: {Wess-Zumino-Witten action and photons from the
  Chiral Magnetic Effect}  (2012)

\bibitem{Fukushima:2010zza}
Fukushima, K., Ruggieri, M.: {Dielectric correction to the Chiral Magnetic
  Effect}.
\newblock Phys. Rev. \textbf{D82}, 054,001 (2010).
\newblock \doi{10.1103/PhysRevD.82.054001}

\bibitem{Fukushima:2007fc}
Fukushima, K., Warringa, H.J.: {Color superconducting matter in a magnetic
  field}.
\newblock Phys. Rev. Lett. \textbf{100}, 032,007 (2008).
\newblock \doi{10.1103/PhysRevLett.100.032007}

\bibitem{Gale:2012rq}
Gale, C., Jeon, S., Schenke, B., Tribedy, P., Venugopalan, R.: {Event-by-event
  anisotropic flow in heavy-ion collisions from combined Yang-Mills and viscous
  fluid dynamics}  (2012)

\bibitem{Gao:2012ix}
Gao, J.H., Liang, Z.T., Pu, S., Wang, Q., Wang, X.N.: {Chiral Anomaly and Local
  Polarization Effect from Quantum Kinetic Approach}  (2012)

\bibitem{Giovannini:1997eg}
Giovannini, M., Shaposhnikov, M.: {Primordial hypermagnetic fields and triangle
  anomaly}.
\newblock Phys. Rev. \textbf{D57}, 2186--2206 (1998).
\newblock \doi{10.1103/PhysRevD.57.2186}

\bibitem{Giovannini:1997gp}
Giovannini, M., Shaposhnikov, M.: {Primordial magnetic fields, anomalous
  isocurvature fluctuations and big bang nucleosynthesis}.
\newblock Phys. Rev. Lett. \textbf{80}, 22--25 (1998).
\newblock \doi{10.1103/PhysRevLett.80.22}

\bibitem{Gorbar:2009bm}
Gorbar, E., Miransky, V., Shovkovy, I.: {Chiral asymmetry of the Fermi surface
  in dense relativistic matter in a magnetic field}.
\newblock Phys. Rev. \textbf{C80}, 032,801 (2009).
\newblock \doi{10.1103/PhysRevC.80.032801}

\bibitem{Gusynin:1994xp}
Gusynin, V., Miransky, V., Shovkovy, I.: {Dimensional reduction and dynamical
  chiral symmetry breaking by a magnetic field in (3+1)-dimensions}.
\newblock Phys. Lett. \textbf{B349}, 477--483 (1995).
\newblock \doi{10.1016/0370-2693(95)00232-A}

\bibitem{Kaiser:2000ck}
Kaiser, R.: {Anomalies and WZW term of two flavor QCD}.
\newblock Phys. Rev. \textbf{D63}, 076,010 (2001).
\newblock \doi{10.1103/PhysRevD.63.076010}

\bibitem{Kalaydzhyan:2012ut}
Kalaydzhyan, T.: {Chiral superfluidity of the quark-gluon plasma}  (2012)

\bibitem{Kalaydzhyan:2011vx}
Kalaydzhyan, T., Kirsch, I.: {Fluid/gravity model for the chiral magnetic
  effect}.
\newblock Phys. Rev. Lett. \textbf{106}, 211,601 (2011).
\newblock \doi{10.1103/PhysRevLett.106.211601}

\bibitem{Kharzeev:2004ey}
Kharzeev, D.: {Parity violation in hot QCD: Why it can happen, and how to look
  for it}.
\newblock Phys. Lett. \textbf{B633}, 260--264 (2006).
\newblock \doi{10.1016/j.physletb.2005.11.075}

\bibitem{Kharzeev:2009fn}
Kharzeev, D.E.: {Topologically induced local P and CP violation in QCD x QED}.
\newblock Annals Phys. \textbf{325}, 205--218 (2010).
\newblock \doi{10.1016/j.aop.2009.11.002}

\bibitem{Kharzeev:2007jp}
Kharzeev, D.E., McLerran, L.D., Warringa, H.J.: {The Effects of topological
  charge change in heavy ion collisions: 'Event by event P and CP violation'}.
\newblock Nucl. Phys. \textbf{A803}, 227--253 (2008).
\newblock \doi{10.1016/j.nuclphysa.2008.02.298}

\bibitem{Kharzeev:2009pj}
Kharzeev, D.E., Warringa, H.J.: {Chiral Magnetic conductivity}.
\newblock Phys. Rev. \textbf{D80}, 034,028 (2009).
\newblock \doi{10.1103/PhysRevD.80.034028}

\bibitem{Kharzeev:2010gd}
Kharzeev, D.E., Yee, H.U.: {Chiral magnetic wave}.
\newblock Phys. Rev. \textbf{D83}, 085,007 (2011).
\newblock \doi{10.1103/PhysRevD.83.085007}

\bibitem{Luscher:1977cw}
Luscher, M.: {SO(4) symmetric solutions of Minkowskian Yang-Mills field
  equations}.
\newblock Phys. Lett. \textbf{B70}, 321 (1977).
\newblock \doi{10.1016/0370-2693(77)90668-2}

\bibitem{Metlitski:2005pr}
Metlitski, M.A., Zhitnitsky, A.R.: {Anomalous axion interactions and
  topological currents in dense matter}.
\newblock Phys. Rev. \textbf{D72}, 045,011 (2005).
\newblock \doi{10.1103/PhysRevD.72.045011}

\bibitem{Nakano:2004cd}
Nakano, E., Tatsumi, T.: {Chiral symmetry and density wave in quark matter}.
\newblock Phys. Rev. \textbf{D71}, 114,006 (2005).
\newblock \doi{10.1103/PhysRevD.71.114006}

\bibitem{Noronha:2007wg}
Noronha, J.L., Shovkovy, I.A.: {Color-flavor locked superconductor in a
  magnetic field}.
\newblock Phys. Rev. \textbf{D76}, 105,030 (2007).
\newblock \doi{10.1103/PhysRevD.76.105030, 10.1103/PhysRevD.86.049901}

\bibitem{Ooguri:2011aa}
Ooguri, H., Oshikawa, M.: {Instability in magnetic materials with dynamical
  axion field}.
\newblock Phys. Rev. Lett. \textbf{108}, 161,803 (2012).
\newblock \doi{10.1103/PhysRevLett.108.161803}

\bibitem{Primakoff:1951pj}
Primakoff, H.: {Photoproduction of neutral mesons in nuclear electric fields
  and the mean life of the neutral meson}.
\newblock Phys. Rev. \textbf{81}, 899 (1951).
\newblock \doi{10.1103/PhysRev.81.899}

\bibitem{Raffelt:1987im}
Raffelt, G., Stodolsky, L.: {Mixing of the Photon with Low Mass Particles}.
\newblock Phys. Rev. \textbf{D37}, 1237 (1988).
\newblock \doi{10.1103/PhysRevD.37.1237}

\bibitem{Rebhan:2009vc}
Rebhan, A., Schmitt, A., Stricker, S.A.: {Anomalies and the chiral magnetic
  effect in the Sakai-Sugimoto model}.
\newblock JHEP \textbf{1001}, 026 (2010).
\newblock \doi{10.1007/JHEP01(2010)026}

\bibitem{Rubakov:2010qi}
Rubakov, V.: {On chiral magnetic effect and holography}  (2010)

\bibitem{Schechter:1977qg}
Schechter, B.M.: {Yang-Mills theory on the hypertorus}.
\newblock Phys. Rev. \textbf{D16}, 3015 (1977).
\newblock \doi{10.1103/PhysRevD.16.3015}

\bibitem{Schon:2000qy}
Schon, V., Thies, M.: {2-D model field theories at finite temperature and
  density}  (2000)

\bibitem{Schwinger:1951nm}
Schwinger, J.S.: {On gauge invariance and vacuum polarization}.
\newblock Phys. Rev. \textbf{82}, 664--679 (1951).
\newblock \doi{10.1103/PhysRev.82.664}

\bibitem{Sikivie:1983ip}
Sikivie, P.: {Experimental Tests of the Invisible Axion}.
\newblock Phys. Rev. Lett. \textbf{51}, 1415 (1983).
\newblock \doi{10.1103/PhysRevLett.51.1415}

\bibitem{Son:2007ny}
Son, D., Stephanov, M.: {Axial anomaly and magnetism of nuclear and quark
  matter}.
\newblock Phys. Rev. \textbf{D77}, 014,021 (2008).
\newblock \doi{10.1103/PhysRevD.77.014021}

\bibitem{Son:2012wh}
Son, D.T., Yamamoto, N.: {Berry curvature, triangle anomalies, and chiral
  magnetic effect in Fermi liquids}  (2012)

\bibitem{Stephanov:2012ki}
Stephanov, M., Yin, Y.: {Chiral Kinetic Theory}  (2012)

\bibitem{Vilenkin:1980fu}
Vilenkin, A.: {Equilibrium parity violating current in a magnetic field}.
\newblock Phys. Rev. \textbf{D22}, 3080--3084 (1980).
\newblock \doi{10.1103/PhysRevD.22.3080}

\bibitem{Yamamoto:2011gk}
Yamamoto, A.: {Chiral magnetic effect in lattice QCD with a chiral chemical
  potential}.
\newblock Phys. Rev. Lett. \textbf{107}, 031,601 (2011).
\newblock \doi{10.1103/PhysRevLett.107.031601}

\bibitem{Yee:2009vw}
Yee, H.U.: {Holographic Chiral Magnetic Conductivity}.
\newblock JHEP \textbf{0911}, 085 (2009).
\newblock \doi{10.1088/1126-6708/2009/11/085}

\bibitem{Zahed:2012yu}
Zahed, I.: {Anomalous chiral Fermi surface}.
\newblock Phys. Rev. Lett. \textbf{109}, 091,603 (2012).
\newblock \doi{10.1103/PhysRevLett.109.091603}

\end{thebibliography}

\end{document}